\documentclass[twocolumn,showpacs,superscriptaddress,preprintnumbers]{revtex4}
\usepackage{graphicx}
\usepackage{dcolumn}
\usepackage{bm}

\begin{document}

\title{Coherence and superconductivity in coupled one-dimensional chains: a
case study of YBa$_{2}$Cu$_{3}$O$_{y}$}
\author{Y.-S. Lee}
\affiliation{Department of Physics, University of California at San Diego, La Jolla,
California 92093-0319 USA}
\author{Kouji Segawa}
\affiliation{Central Research Institute of Electric Power Industry, Komae, Tokyo
201-8511, Japan}
\author{Yoichi Ando}
\affiliation{Central Research Institute of Electric Power Industry, Komae, Tokyo
201-8511, Japan}
\author{D. N. Basov}
\affiliation{Department of Physics, University of California at San Diego, La Jolla,
California 92093-0319 USA}
\date{\today }

\begin{abstract}
We report the infrared (IR) response of Cu-O chains in the high-$T_{c}$
superconductor YBa$_{2}$Cu$_{3}$O$_{y}$ over the doping range spanning $%
y=6.28-6.75$. We find evidence for a power law scaling at mid-IR frequencies
consistent with predictions for Tomonaga-Luttinger liquid, thus supporting
the notion of one-dimensional transport in the chains. We analyze the role
of coupling to the CuO$_{2}$ planes in establishing metallicity and
superconductivity in disordered chain fragments.
\end{abstract}

\maketitle

The appeals of one-dimensional (1D) electronic systems are many. Indeed,
foundational concepts of condensed matter physics are revised in the one
dimension \cite{Dressel-review}. The conventional quasiparticle description
breaks down in 1D solids and the spin-charge separation concept needs to be
invoked to understand excitations. An experimental exploration of 1D physics
spans a diversity of systems ranging from conducting molecules to
unidirectional charge-ordered regions (stripes) in high-$T_{c}$
superconductors and quantum Hall structures. Many aspects of 1D physics
including the formation of the correlation gap in the electrodynamic
response \cite{Giamarchi97} are well understood in part owing to the advent
of exactly solvable models \cite{Tsvelik01}. An arrangement of coupled 1D
conductors is envisioned as a paradigm to explain unconventional properties
at higher dimensions specifically in the context to the problem of high-$%
T_{c}$ superconductivity. While the principal building blocks of cuprate
superconductors are two-dimensional CuO$_{2}$ planes, a prototypical family
of high-$T_{c}$ materials YBa$_{2}$Cu$_{3}$O$_{y}$ (YBCO) does also possess
1D elements -- the so-called CuO chains. In (nearly) stoichiometric $y=6.95$
and YBa$_{2}$Cu$_{4}$O$_{8}$ compounds chains are conducting and also show
substantial contribution to the superfluid density \cite{Ando02,Basov95}.

Here we investigate the emergence of coherent response and superconductivity
in the electrodynamics of the chains in underdoped YBCO compounds ($%
y=6.28-6.75$). In these materials chain fragments extend over distances 15 -
400 \AA\ (depending on the doping), which is insufficient to maintain dc
currents across macroscopic specimens. Our infrared (IR) experiments show
that when the chain fragment length exceeds a critical value their coupling
via neighboring CuO$_{2}$ planes enables direct contribution of the CuO
chains to the dc transport and superconductivity. At frequencies above the
correlation gap (50 -- 200 meV, another doping-dependent parameter) the
electrodynamics of chain fragments reveals universal scaling behavior
previously found in other classes of 1D conductors \cite%
{Schwartz98,Vescoli98, Takenaka00} in accord with predictions for the
Tomonaga-Luttinger (TL) liquid. We emphasize that the interplay between
superconductivity and correlated insulating state is difficult to
investigate in more conventional 1D systems. In this regard YBCO compounds
constitute an excellent test bed for a survey of the issues pertaining to
superconductivity of 1D coupled conductors as will be elaborated below.

\begin{figure}[tbp]
\includegraphics[width=0.48\textwidth]{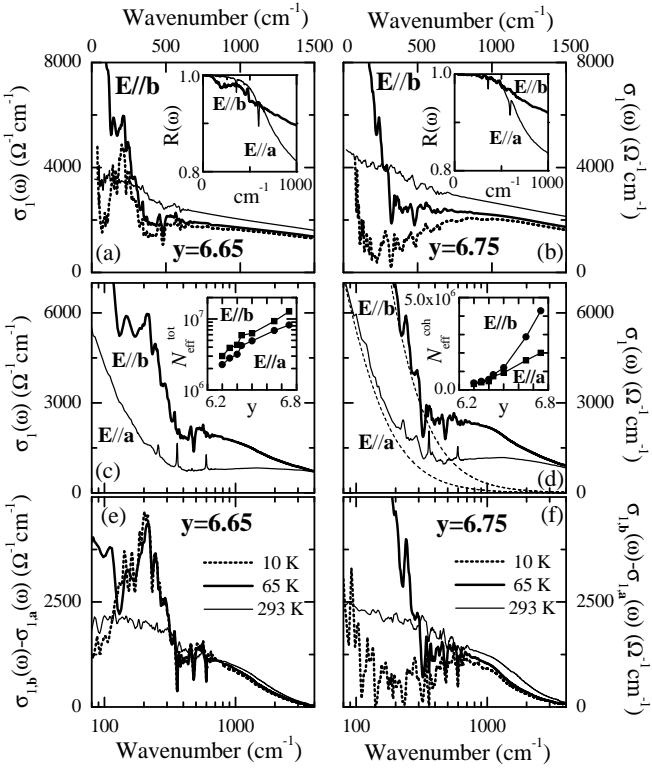}
\caption{Spectra of the real part of the conductivity for $y$ = 6.65 (left
panels) and $y$ = 6.75 crystals (right panels). On the top we show the $b$%
-axis data with raw reflectivity results at 10 K displayed in the insets.
Middle panels depict the $a$- and $b$-axis spectra at 65 K, with doping
dependence of the total intra-gap spectral weight $N_{\text{eff}}^{tot}$
(left inset) and the coherent part of the spectral weight $N_{\text{eff}%
}^{coh}$ plotted (right inset). In (d), the dotted lines represent the Drude
fitting results for coherent modes. The fitted dc resistivity and scattering
rate are 60 (35) $\protect\mu \Omega $cm and 75 (105) cm$^{-1}$ for the $a$ (%
$b$) axis, respectively. Bottom panels (e, f) show the difference in
conductivity $\protect\sigma ^{ch}(\protect\omega )=\protect\sigma _{1,b}(%
\protect\omega )-\protect\sigma _{1,a}(\protect\omega )$ attributable to the
response of the chains as described in the text. One finds a resonance
structure centered at $\sim $ 200 cm$^{-1}$ in the low-$T$ $b$-axis data
(panels a,c, and e) for the $y=6.65$ sample.}
\end{figure}

We investigated detwinned YBCO single crystals with oxygen content $y$ =
6.28, 6.30, 6.35, 6.40 ($T_{c}$ $\sim $ 2 K), 6.43 ($T_{c}$ $\sim $ 13 K),
6.50 ($T_{c}$ $\sim $ 31 K), 6.55 ($T_{c}$ $\sim $ 50 K), 6.65 ($T_{c}$ $%
\sim $ 60 K), and 6.75 ($T_{c}$ $\sim $ 65 K) grown by a conventional flux
method and detwinned under uniaxial pressure \cite{Segawa01}. Annealing
under the uniaxial pressure also aligns chain fragments along the $b$-axis
in non-superconducting YBCO ($y$ = 6.28 - 6.35). Reflectivity spectra $%
R(\omega )$ at nearly normal incidence were measured with polarized light at
frequencies from 20 to 48,000 cm$^{\text{-}1}$ and at temperatures from 10
to 293 K. The complex optical conductivity spectra, $\widetilde{\sigma }%
(\omega )=\sigma _{1}(\omega )+i\sigma _{2}(\omega )$, were obtained from
the measured $R(\omega )$, using the Kramers-Kronig (KK) transformation. The
KK-derived results are consistent with those obtained by ellipsometry. Note
that the chains in the YBCO system extend along the $b$-axis of the crystal
and therefore do not contribute to the conductivity probed in the
polarization \textbf{E }$\parallel a$. On the contrary, both the CuO$_{2}$
planes and the 1D Cu-O chains jointly contribute to the $b$-axis spectra.

We first review the key features of the electromagnetic response of
underdoped YBCO. Below the charge transfer excitation near 14,000 cm$^{-1}$
one can identify two distinct absorption channels in the dissipative part of
the conductivity. At the lowest frequencies ($\omega <600$ cm$^{-1}$) one
recognizes a coherent contribution that has the Drude-like form. In
disordered samples where localization effects are important, the Drude mode
may evolve into a resonance centered at non-zero frequency. This particular
transformation is especially clear in the $b$-axis conductivity of the $%
y=6.65$ crystal showing a resonance at 200 cm$^{-1}$ at low $T$ \ [Fig.
1(a)]. This low frequency mode has to be contrasted with the smooth spectra
of the $y=6.75$ phase in Fig. 1(b). The coherent contribution to the
conductivity is strongly $T$-dependent and the changes of $\sigma (\omega
,T) $ are directly related to transport data. Apart from the coherent low-$%
\omega $ mode a broad absorption structure is clearly seen in the mid-IR
conductivity. The mid-IR absorption does not show significant modifications
with $T$. With increased doping both features grow in strength and can no
longer be distinguished in optimally doped phases.

\begin{figure}[tbp]
\includegraphics[width=0.32\textwidth]{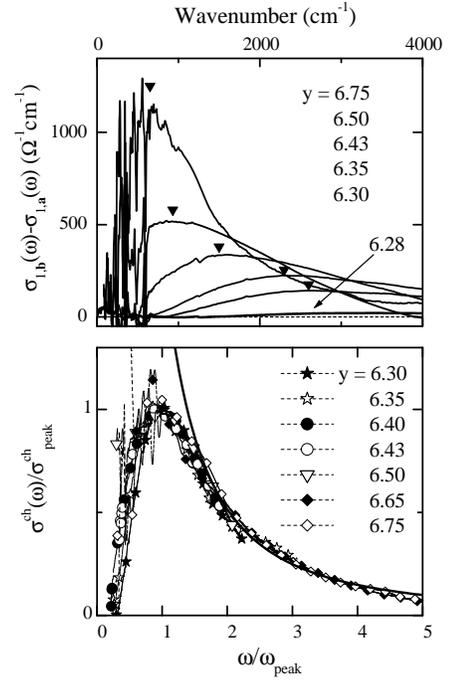}
\caption{(a) Spectra of $\protect\sigma ^{ch}(\protect\omega )=\protect%
\sigma _{1,b}(\protect\omega )-\protect\sigma _{1,a}(\protect\omega )$ at
lowest temperatures for a series of YBCO crystals. For $y=6.75$, the
coherent mode is not visible because its weight is tranferred to
superconducting $\protect\delta $-peak at $\protect\omega =0$. (b) $\protect%
\sigma ^{ch}(\protect\omega )/\protect\sigma _{\text{peak}}^{ch}$ with $%
\protect\omega /\protect\omega _{\text{peak}}$ as the abscissa. For clarity,
the sharp phonon structures are removed from the data in the bottom panel.
The solid line represents $\protect\omega ^{-\protect\alpha }$-dependence
with $\protect\alpha =1.6$.}
\end{figure}

We now turn to the analysis of the effects in the electromagnetic response
attributable to Cu-O chain fragments. It is instructive to consider various
ways in which the chains influence charge dynamics by introducing the
electronic spectral weight $N_{\text{eff}}(\omega )=\int_{0}^{\omega }\sigma
_{1}(\omega ^{\prime })d\omega ^{\prime }$. In the inset of Fig.~1(c) and
1(d), we show the doping dependence of the total intra-gap spectral weight $%
N_{\text{eff}}^{tot}=N_{\text{eff}}(\omega =10,000$ cm$^{-1})$ and that of
the coherent component alone $N_{\text{eff}}^{coh}=N_{\text{eff}}(\omega
=600 $ cm $^{-1})$ separately for the $a$-axis and $b$-axis data. A quick
inspection of the plots shows that the chains increase the total spectral
weight registered in the $b$-axis conductivity by as much as $\sim $ 40 \%
in all underdoped samples in accord with the earlier data \cite{Cooper93}.
The doping dependence of the coherent spectral weight is different: $N_{%
\text{eff }}^{coh}$ is nearly isotropic for $y\leq 6.50$, but at higher
dopings $N_{\text{eff}}^{coh}$ acquires anisotropy that is progressively
increasing with $y$. An enhancement of the $N_{\text{eff}}^{coh}$ in the $b$%
-axis response may be attributed to the \textit{direct} contribution of the
Cu-O chains to the far-IR conductivity \cite{imprinting-footnote}. We stress
that the anisotropy of the coherent contribution is observed only when the
doping exceeds a critical value above $y=6.50$ for this set of samples.
X-ray, NMR, and optical experiments convincingly show that the length of the
chain segments is progressively increasing with doping \cite%
{liang00,liang02,yslee04}. Given this trend, we therefore conclude that a
minimum length of chains is required to trigger the coherent component \cite%
{comment-chainlength}.

It is useful to supplement the analysis of the anisotropy of the electronic
spectral weight with the examination of $\sigma ^{ch}(\omega )=\sigma
_{1,b}(\omega )-\sigma _{1,a}(\omega )$. Assuming that the chains produce a
parallel conductivity channel in YBCO, the spectra of $\sigma ^{ch}(\omega )$
are attributable to the response of the chains \cite{Cooper93}. For $y\leq
6.50$, in accord with the $N_{\text{eff}}$ analysis discussed above, the $%
\sigma ^{ch}(\omega )$ reveal no significant electronic contribution below
600 cm$^{-1}$ [top panel of Fig. 2(a)]. The dominant chain contribution to
optical conductivity in these samples is an asymmetric mid-IR mode revealing
considerable softening with the increase of $y$. The similar structure at
lower $\omega $ can be identified in the $\sigma ^{ch}(\omega )$ data for
both the $y=6.65$ and 6.75 crystals. In addition, we have been able to
detect significant low-frequency conductivity in the latter compounds,
uncovering interesting $T$-dependence [bottom panel of Fig. 1]. The low-$%
\omega $ coherent response of the chains above $T_{c}$ in the $y=6.75$
sample leads to $\sigma ^{ch}(\omega )$ that is monotonically increasing as $%
\omega \rightarrow 0$, whereas the dominant contribution to the low $T$ $%
\sigma ^{ch}(\omega )$ for the $y=6.65$ compound is a narrow resonance at
200 cm$^{-1}$.

We now focus on the analysis of the mid-IR mode in the $\sigma ^{ch}(\omega
) $ spectra. The bottom panel of Fig.~2 shows the scaling behavior of this
mode. The diagram demonstrates that all spectra collapse on a single curve
following $\omega /\omega _{\text{peak}}$ protocol for the abscissa and $%
\sigma ^{ch}(\omega )$/$\sigma _{\text{peak}}^{ch}$ renormalization for the
vertical axis. Here, $\omega _{\text{peak}}$ and $\sigma _{\text{peak}}^{ch}$
represent the position and the height of the mid-IR mode, respectively. The
scaling behavior persists up to $\sim $ 0.5 eV and is obeyed for the
variation of $\omega _{\text{peak}}$ from 650 cm$^{-1}$ for $y=6.75$ to
2,500 cm$^{-1}$ for $y=6.30$ \cite{comment-Eg}, attesting to the
significance of this result. Although the gross features of the $a$-axis
conductivity of YBCO and other cuprates are similar to those seen in the $%
\sigma ^{ch}(\omega )$ spectra, the scaling trend is not identified in the $%
a $-axis. We therefore conclude that physics underlying the scaling law is
intimately related to the 1D character of the electronic transport in the
chains. This conjecture is supported by theoretical analyses of the optical
conductivity of 1D correlated insulators within the TL theory \cite%
{Giamarchi97,Tsvelik01}. These models prescribe the emergence of a
correlation gap $E_{g}$ in a 1D system near commensurate fillings leading to
a strong asymmetric resonance in the $\sigma (\omega )$ at $\omega \simeq
E_{g}$. When the filling deviates from the commensurability due to doping, a
coherent mode is expected to emerge within the correlation gap. Hence, a
salient feature of the metal-insulator transition in a correlated 1D system
is the development of a \textit{two-component} form of the optical
conductivity. Remarkably, the spectra of $\sigma (\omega )$ for $\omega
>E_{g}$ are predicted to follow the same power law $\sigma (\omega )\propto
\omega ^{-\alpha }$ behavior on both metallic and insulating sides of the
transition. Thus the gross features of the electromagnetic response
attributable to 1D Cu-O chains as well as the evolution of these trends with
doping are consistent with the predictions of the TL theory.

Hallmarks of one-dimensionality uncovered by the above analysis of the $%
\sigma ^{ch}(\omega )$ spectra motivate us to compare electrodynamics of the
Cu-O chains with the behavior of more conventional 1D conductors complying
with the predictions of the TL model. The observed power law with $\alpha
=1.6$ in Fig. 2 is distinct from the $\omega ^{-3}$ response expected for a
band insulator \cite{Giamarchi97}, but is close to $\alpha =1.3$ seen in 1D
Bechgaard salts \cite{Schwartz98,Vescoli98}. The range of $E_{g}$ values in
YBCO is comparable to that of the Bechgaard salts as well \cite%
{Schwartz98,Vescoli98}. An important feature of the TL model is that
universal 1D scaling characteristics are expected to be found only at
frequencies above $E_{g}$ whereas at lower energies ($\omega <E_{g}$) a
system of coupled 1D conductors shows 2D or 3D behavior \cite%
{Giamarchi97,carlson02}. This dimensional crossover is seen in Bechgaard
salts \cite{Giamarchi97,Schwartz98,Vescoli98} and also can be identified in
the $\sigma ^{ch}(\omega )$ data for YBCO crystals with $y$ > 6.50 revealing
a prominent coherent contribution. Notably, the low-energy properties of the
TL liquid are non-universal since this behavior critically depends on the
details of the inter-chain coupling. An important aspect of the coupling in
YBCO is a close proximity of the chains to the CuO$_{2}$ planes implying
that the hybridization between the chain and plane states has to be taken
into account \cite{atkinson99}. This strong coupling is likely to be
responsible for the pronounced spectral weight of the coherent mode relative
to the total weight in the $\sigma ^{ch}(\omega )$ spectra. Indeed, we find
that for YBCO $N_{\text{eff}}^{coh}/N_{\text{eff}}^{tot}\sim 40-50$ \%
whereas in Bechgaard salts this ratio is as small as 1-2 \% \cite%
{Schwartz98,Vescoli98}. Given the strong chain-to-plane coupling when CuO$%
_{2}$ layers may act as Ohmic heat baths for charge degrees of freedom \cite%
{morr01}, it is not surprising that the power laws close to the Fermi liquid
pattern seen in the $T$ dependence of the chain resistivity $\rho ^{ch}(T)$\ 
$[=1/(\sigma _{b}-\sigma _{a})]$ \cite{Gagnon94} are different from those
observed in 1D organic conductors.

\begin{figure}[tbp]
\includegraphics[width=0.38\textwidth]{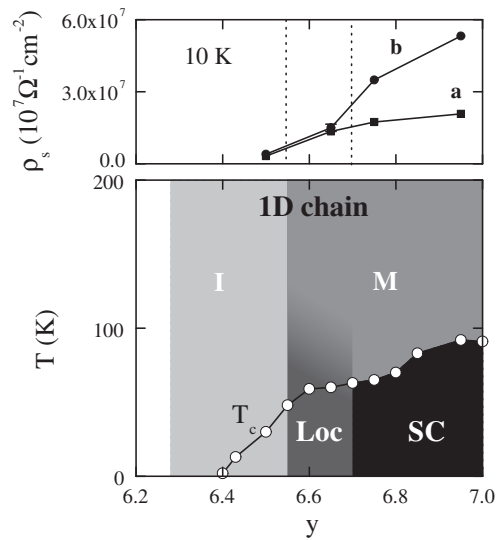}
\caption{Top panel: the doping dependence of the superfluid density $\protect%
\rho _{s}$ at $y$ = 6.50 - 6.95 \protect\cite{Basov95} separately for $a$%
-axis and $b$-axis data. Bottom panel: phase diagram displaying
characteristic regimes in the response of the CuO chains. Here I stands for
insulating, LOC for localized metal, M for Drude metal, and SC for
superconducting. }
\end{figure}

We now turn to the analysis of the transformation of the spectra in the
superconducting state. A spectroscopic signature of superconductivity is the
transfer of the electronic spectral weight from finite frequencies to
superconducting $\delta $-function at $\omega =0$. The strength of the $%
\delta (0)$-peak referred to as the superfluid density $\rho _{s}$ can be
readily evaluated from the so-called missing area in the conductivity data,
following the Ferrel-Glover-Tinkham sum rule: $\rho _{s}=8\int d\omega
\lbrack \sigma _{1}(\omega ,T>T_{c})-\sigma _{1}(\omega ,T\ll T_{c})]$ or
from the imaginary part of the conductivity \cite{Dordevic02}. Data in top
panel of Fig. 3 uncover nearly isotropic $\rho _{s}$ in the $y=6.65$
crystal, whereas for $y=6.75$ the superconducting condensate is strongly
enhanced in the $b$-axis data. This contrasting behavior of the two crystals
can be traced back to distinctions in the low-frequency coherent
contribution to $\sigma ^{ch}(\omega )$ at $T\simeq T_{c}$. Indeed, in the
the $y=6.75$ crystal one finds a Drude-like form of the low-$\omega $
conductivity whereas in the $y=6.65$ sample the coherent contribution
produces a finite energy resonance centered at 200 cm$^{-1}$. Similar
absorption structures are commonly found in the optical conductivity data
for low-dimensional disordered conductors, and can be attributed to the weak
localization induced by defects or disorders \cite{localization}.
Apparently, bound carries forming a 200 cm$^{-1}$ resonance are excluded
from contributing to $\rho _{s}$. We also plotted the $y$-dependent $\rho
_{s}$ from $y=6.50$ to 6.95 \cite{Basov95}. Data in Fig. 3 show that $\rho
_{s}$ along the $b$-axis starts to exceed that probed along the $a$-axis for 
$y>6.65$, coincident with the disappearance of the localization mode at 200
cm$^{-1}$. Evidently, the delocalization of the bound carriers yields an
additional contribution to the superfluid density resulting in anisotropic $%
\rho _{s}$ in YBCO.

The experiments detailed above uncover several new facets of the response of
1-D conducting elements in the environment of doped MH insulators. The
scaling behavior of $\sigma ^{ch}(\omega )$ reaffirms the applicability of
the TL description to the response of the Cu-O chains segments in YBCO. Our
measurements indicate that the chain segment length has to exceed a minimum
value (15 - 20 \AA ) to trigger the universal power law in $\sigma (\omega )$
data \cite{liang02,yslee04}. Indeed, no readily identifiable resonance is
found in the data for the $y$ = 6.28 crystal which has an average length of
Cu-O segments below this cut-off [Fig. 2(a)]. The proximity to highly
conducting CuO$_{2}$ planes defines an entirely different character of the
coherent mode in the $\sigma ^{ch}(\omega )$ spectra. Specifically, the
plane-mediated processes dramatically enhances the strength of the coherent
mode compared to modes generated by the direct inter-chain hopping in
conventional 1-D systems. Once the chain fragment length exceeds the
critical value and the separation between these fragments is reduced, the $%
\sigma ^{ch}(\omega )$ spectra reveal the Drude-like metallic behavior which
would be impossible in a system of isolated disordered chains. In between
these two regions one finds a territory marked as `localized metal' where
chains do contribute to the electronic conductivity but not to the
superfluid density. The superconducting condensate emerges as soon as the
inter-chain coupling is strong enough to support conventional Drude-like
behavior at $T\simeq T_{c}$. The particular dopings corresponding to the
transitions from the insulator to the localized metal and to metal in $%
\sigma ^{ch}(\omega )$ are likely to be sensitive to details of sample
preparation. Due to the effective hybridization between the plane and the
chain states, the superconductivity residing in the chain might have
three-dimensional character rather than one-dimensional. Recall that the
usual description of superconductivity in the chains via a proximity
coupling to the planes has difficulties in accounting for the identical $T$
dependence of the superfluid density measured along the $a$- and $b$-axis in
YBCO \cite{atkinson95}. In this context it is interesting to note that
superconducting ground state is one of possible ordered phases predicted to
arise from inter-chain interaction within the TL model \cite{Giamarchi97}.
This corollary of the TL scenario warrants further analysis of
superconducting response of the chains using a variety of relevant probes.
The doping dependent trend in chains is reminiscent of the
superconductor-insulator transition in disordered CuO$_{2}$ planes \cite%
{Basov98}.

This research was supported by the US DOE.


\begin{thebibliography}{99}
\bibitem{comment-chainlength} It is known that details of sample preparation
may strongly influence the chain length and its estimates in the literature
for $y$ $\sim 6.50$ range from 200 \AA\ \cite{liang02} to 400 \AA\ \cite%
{liang00}.
\end{thebibliography}
\end{document}